\documentclass{article}
\usepackage{geometry}
\usepackage{amsmath,amssymb}
\usepackage{graphicx}
\graphicspath{ {./figs/} }
\usepackage{dcolumn}
\usepackage{bm}
\usepackage{hyperref}
\usepackage{enumitem}
\usepackage{tikz}
\usepackage{pgfplots}
\pgfplotsset{compat=1.3}
\usepackage{caption}
\usepackage{subcaption}
\usepackage{float}
\usepackage{stanli}

\usepackage[style=nature,doi=false,isbn=false,url=false,eprint=false]{biblatex}
\usepackage{lineno}

\usepackage[utf8]{inputenc}
\usepackage{graphicx}
\usepackage{amsmath}
\usepackage{gensymb}
\usepackage{multirow}
\usepackage{cleveref}
\usepackage{caption}
\usepackage{float}
\usepackage{xcolor}
\usepackage{authblk}

\definecolor{verydarkgreen}{HTML}{004D40}

\begin{document}


\title{The deflection limit of slab-like topologically interlocked structures}

\author[1]{Silvan Ullmann}
\author[1]{David S. Kammer}
\author[1]{Shai Feldfogel\footnote{corresponding author: +41 44 633 05 44, sfeldfogel@ethz.ch}}

\affil[1]{Institute for Building Materials, ETH Zurich, Switzerland}

\maketitle

\begin{abstract}
Topologically Interlocking Structures (TIS) are structural assemblies that achieve stability and carrying capacity through the geometric arrangement of interlocking blocks, relying solely on normal and friction forces for load transfer. Unlike beam-like TIS, whose deflection never exceeds the height of the blocks, the deflection of slab-like TIS often does. Yet, the upper limit of deflection of slab-like TIS, a key parameter defining their loading energy capacity, remains unexplored. Here, we establish a theoretical upper bound for the deflection capacity of slab-like TIS and outline a systematic design strategy to approach this upper bound. This strategy is based on engineering the contact interfaces such that the non-central blocks are more engaged in the structural response, leading to a more global and holistic deformation mode with higher deflections. We demonstrate the application of this strategy in a numerical case study on a typical slab-like TIS. We show that it leads to a 350$\%$ increase in deflection, yielding a value closer to the upper bound than previously reported in the literature. We also show that the resulting deflection mode engages all the blocks equally, avoids localized sliding modes, and resembles that of monolithic equivalents. Lastly, we show that the strategy not only maximizes the assembly's deflection capacity but also its loading energy capacity.
\end{abstract}

\section{Introduction}
\label{sec:introduction}

Topologically Interlocking Structures (TIS) are structural assemblies made of specially shaped and arranged unbonded building blocks. Unlike ordinary structures, which derive structural integrity and continuity from their monolithic nature (e.g.,  reinforced concrete), or from structural connections used to connect their constituent members (e.g., steel structures), the structural integrity of TIS relies entirely on contact and friction forces that develop at the block interfaces \cite{Dyskin2003a, Dyskin2003b}.

One of the most attractive properties of TIS, considering they are built from brittle materials with low toughness, is their ability to absorb and dissipate energy through hinging and frictional sliding between the blocks \cite{Dyskin2003c, Schaare2008, Schaare2009, Khandelwal2012, Krause2012, Feng2015, Khandelwal2015, Molotnikov2015, Mirkhalaf2018, Short2019, RezaeeJavan2020}. This quality is directly related to the deflections that TIS can develop, which highlights the importance and relevance of the deflection capacity and limit of TIS. \par\medskip

\begin{figure}[H]
    \includegraphics[width=0.6\textwidth]{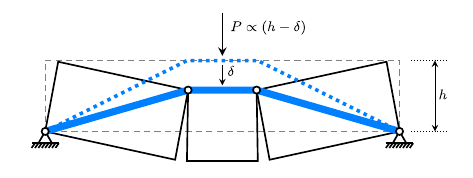} 
    \centering
    \caption{Thrust line analogy for load transmission along a row of TIS blocks.}
    \label{fig:thrustline}
\end{figure}

In beam-like TIS structures, schematically represented in Fig.~\ref{fig:thrustline}, the deflection $\delta$ never exceeds the height of the beam $h$ \cite{Dalaq2019, Dalaq2020, Koureas2022, Koureas2023, Laudage2023}. 
This is because the load $P$ is proportional to the effective static depth $h-\delta$, in accordance with the thrust line model for TIS \cite{Khandelwal2012, Khandelwal2013, Short2019}, see Fig.~\ref{fig:thrustline}.

Contrary to beam-like TIS, the maximal deflection in slab-like TIS panels often exceeds the panel's thickness \cite{Khor2002, Brugger2009, Schaare2009, Krause2012, Khandelwal2012, Feng2015, Molotnikov2015, Khandelwal2015, Djumas2017, RezaeeJavan2020, Williams2021, Kim2021, Feldfogel2022, Feldfogel2023}. This raises the following questions, which are at the focus of this study: What is the mechanism that allows TIS slabs to develop deflections larger than the slab thickness? Is there a limit to the deflection of slab-like TIS, and, if so, what is it? Is there a systematic way to design slab-like TIS such that they approach this deflection limit? How does the deflection limit relate to their energy absorption/dissipation capacity, a key design criterion for some applications?

Several aspects concerning the ability of slab-like TIS to absorb and dissipate energy have been considered in the literature. This includes how the loading energy can be controlled by active and adaptive external constraints \cite{Khandelwal2015}, how it depends on the number and the geometry of the blocks \cite{Mirkhalaf2018, Weizmann2021, Short2019}, how it is scales with Young's modulus and the friction coefficient \cite{Feldfogel2023}, and how inserting soft interfaces between the blocks can massively increase the deflections \cite{RezaeeJavan2020}. Nevertheless, to our knowledge, the deflection limit of slab-like TIS and its implications on the energy absorption capacity remain unexplored. The objective of this study is therefore to establish a theoretical upper bound for the deflection of slab-like TIS, to propose a systematic strategy to approach this theoretical limit, and to demonstrate the link between approaching the deflection limit and locally maximizing the loading energy capacity. \par\medskip

Next, in Sec.~\ref{sec:methodology}, we describe and illustrate the theoretical upper bound for the deflection of slab-like TIS considering an approximate and hypothetical deformation mode. Based on this upper bound and on the typical structural response of TIS slabs, we then outline a general strategy to maximize the deflection limit. Lastly, we briefly discuss the Level-Set-Discrete-Element-Method (LS-DEM), the computational tool with which we demonstrate a practical application of this strategy.
In Sec.~\ref{sec:results}, we apply the strategy to maximize the deflection capacity to one of the slabs studied both experimentally and numerically in \cite{Mirkhalaf2018, Feldfogel2023}, and discuss the practical significance of the results for TIS design.
Finally, in Sec.~\ref{sec:conclusions} we conclude the study and outline directions for future research.

\section{Methodology}
\label{sec:methodology}

\subsection{The reason why the deflection in TIS slabs can be larger than the thickness} \label{sec:why_slabs_can_deflect_more_than_h}

\begin{figure}[H]
    \includegraphics[width=0.8\textwidth]{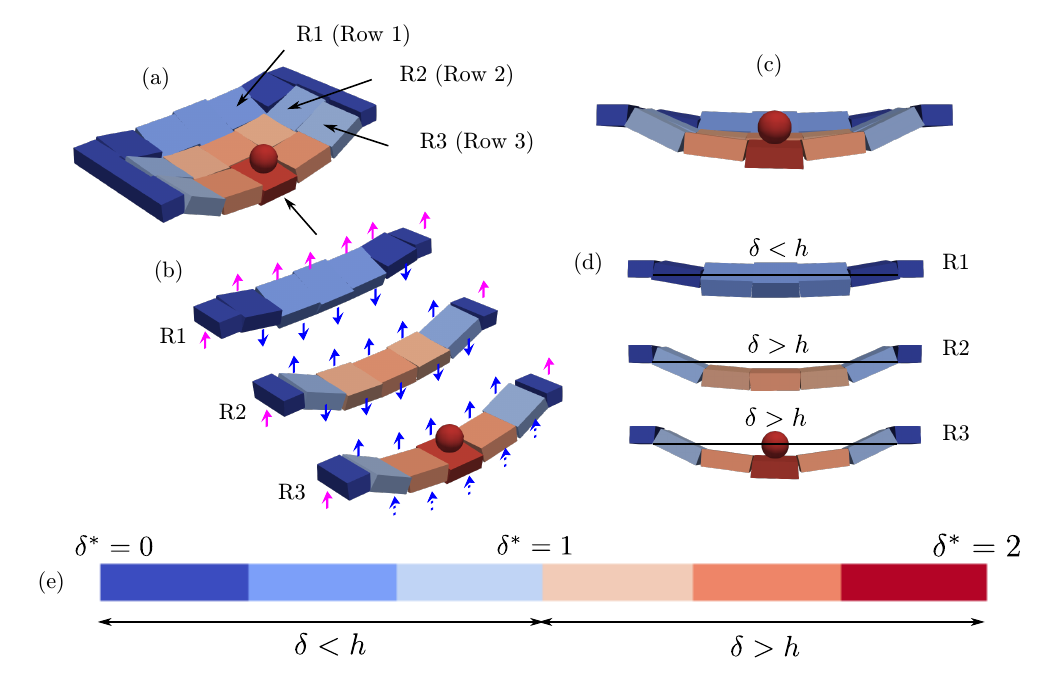} 
    \centering
    \caption{Illustration of the chaining mechanism through which the deflection $\delta$ in slab-like TIS can exceed the slab thickness $h$. (a)~Isometric view of the first three rows of blocks in a 5 x 5 TIS. (b)~Schematic representation of the vertical contact forces between the three rows and the support reaction forces, indicated with blue and magenta arrows, respectively. The dashed arrows indicate forces applied by Row 4 (not shown in the figure). (c)~View on the three rows from the direction indicated by the black arrow in (a). (d)~Deflected shape of each of the three rows. (e)~Normalized deflection color bar for $\delta^*=\delta / h$.}
    \label{fig:chaining_effect}
\end{figure}

Contrary to TIS beams, which transmit loads to the supports along a single path, in slab-like TIS panels, the loads are transmitted to the peripheral line supports through multiple rows of blocks \cite{Khandelwal2012,Khandelwal2013,Short2019}. 
The interaction between rows of blocks allows slab-like TIS to have deflections larger than the thickness, as illustrated in Fig.~\ref{fig:chaining_effect}.
When a row of blocks, typically the central one, is subjected to a downward load, part of it is carried to the supports through the thrust-line mechanism described in Fig.~\ref{fig:thrustline}.
Another part is distributed to parallel adjacent rows of blocks through frictional forces, see Fig.~\ref{fig:chaining_effect}(b). The resulting upward reaction on the loaded row allows it to continue carrying loads even when its deflection is larger than the height of the blocks, see Fig.~\ref{fig:chaining_effect}(b) and (d). This "chaining" force transmission mechanism between rows of blocks can continue, but not indefinitely. The row nearest to the support cannot deflect more than $h$ if it is to receive the upward reaction from the fixed boundary blocks, shown in  Fig.~\ref{fig:chaining_effect}(d). Provided that this edge row holds said condition, the internal rows can rely on the described adjacent-row-forces in addition to the thrust-line carrying mechanism. 

The chaining mechanism explains the recurring observation in experiments and simulations alike that the ultimate deflection in TIS slabs exceeds the thickness of the blocks \cite{Khor2002, Brugger2009, Schaare2009, Krause2012, Khandelwal2012, Feng2015, Molotnikov2015, Khandelwal2015, Djumas2017, RezaeeJavan2020, Williams2021, Kim2021, Feldfogel2022, Feldfogel2023}.
Noting that the forces transmitted from the central row to the supports cause a twisting moment that tends to rotate rows 1 and 2 about their axes, see Fig. \ref{fig:chaining_effect}(b), this mechanism also explains why the blocks rotate as they typically do.

\subsection{The upper bound of the deflection capacity of slab-like TIS} \label{sec:upper_bound}
\begin{figure}[H]
    \centering
    \includegraphics[width=0.85\linewidth]{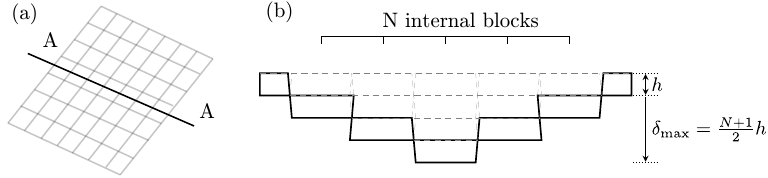}
    \caption{The deformed configuration on which the upper bound expression for the deflection capacity is derived. (a) A schematic of a central cross-section in a slab-like TIS. (b) The deformed configuration of a central row (section A-A) which defines the maximal possible deflection $\delta_\text{max}$.}
    \label{fig:theretical_limit_mechanism}
\end{figure}

The discussion in the previous section raises an obvious follow-up question: What is the upper limit of the deflection of slab-like TIS?
We now propose an upper bound for the normalized deflection $\delta^*=\delta / h$ of slab-like TIS, based on their geometry and typical deformation modes. 
It is inspired by the observation that the deformation mode of slab-like TIS under high loads is one where the loaded block (typically the central one) and its neighbors contribute to the global deformation much more than the peripheral ones. 
Specifically, realizing that it is this minor engagement of peripheral blocks that limits the ability to develop large deflections, we chose  an idealized deformation mode with equal participation of all the blocks as the basis for the upper bound. 
This mode is depicted in Fig. \ref{fig:theretical_limit_mechanism}(b) for a centrally loaded square panel with $N$~x~$N$ internal blocks. 
It is entirely defined by the deformed configuration of the central row with minimal possible contact area between the blocks.
The blocks in non-central rows are free to assume different combinations of stick, sliding, and rotations, as in Fig.~\ref{fig:chaining_effect}, provided they remain in contact.

From the deformation mode in Fig. \ref{fig:theretical_limit_mechanism}(b), we define the maximal possible normalized deflection as: 
\begin{equation}
    \delta_\text{max}^* = \frac{N+1}{2} \label{eq:deflection_upper_bound}   
\end{equation}

The proportionality between the deflection and the number of blocks in Eq.~\ref{eq:deflection_upper_bound} also appears in Short \& Siegmund \cite{Short2019}.
Their expression (Eq. 16) defines the deflection at which the load becomes zero for a given panel with NxN blocks, and, curiously, it yields a deflection equal to exactly half of the upper bound we propose\footnote{The reason for this is that, both here and in \cite{Short2019}, it is assumed that the transverse displacements of the central row vary linearly from the support to the mid-span. This means that the farther away from the center the row where $\delta=h$ is located, the larger the deflection at the center will be. And, since in our case this row is twice as far from the center compared to \cite{Short2019}, our deflection is twice as large.}. 

\begin{figure}
    \centering
    \includegraphics[width=0.8\linewidth]{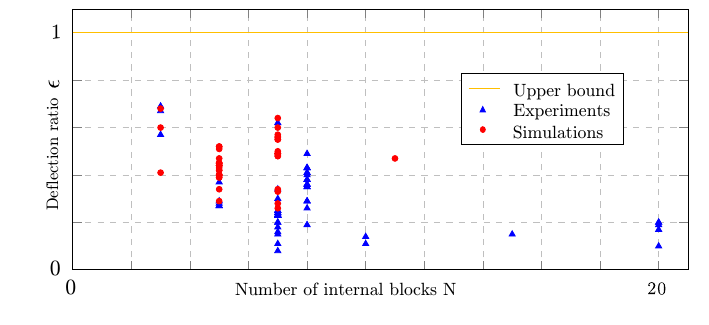}
    \caption{Deflection ratio $\epsilon$ for experimental and numerical results reported in the literature \cite{Khor2002, Dyskin2003b, Dyskin2003c, Brugger2009, Schaare2009, Krause2012, Mather2012, Khandelwal2012, Feng2015, Molotnikov2015, Khandelwal2015, Djumas2017, Mirkhalaf2018, RezaeeJavan2020, Williams2021, Kim2021, Feldfogel2022, Feldfogel2023}. All results fall within the proposed upper bound $\epsilon=1$.}
    \label{fig:ductility_literature}
\end{figure}

To examine the proposed upper-bound, we compared it with all the relevant results found in the literature for slab-like TIS\footnote{In Rezaee Javan et al. \cite{RezaeeJavan2020} a hybrid assembly made of stiff blocks with soft interlayers inserted in-between was tested. As this configuration is qualitatively different from standard ones, which do not have soft interlayers, it is not included in the discussion} \cite{Khor2002, Dyskin2003b, Dyskin2003c, Brugger2009, Schaare2009, Krause2012, Mather2012, Khandelwal2012, Feng2015, Molotnikov2015, Khandelwal2015, Djumas2017, Mirkhalaf2018, RezaeeJavan2020, Williams2021, Kim2021, Feldfogel2022, Feldfogel2023}. These examples include a wide range of experimental and computational results from slab-like TIS with a wide range of materials, block geometries, etc. 
To compare assemblies with different thicknesses and number of blocks, we define the deflection ratio $\epsilon$ as:
\begin{equation}
    \epsilon = \frac{\delta^*}{\delta_\text{max}^*} = \frac{2 \cdot \delta}{(N+1) \cdot h} \leq 1    \label{eq:deflection_ratio}
\end{equation}
and use it as the comparative measure.
Note that a case with $\epsilon=1$ means that it reached the upper bound $\delta^{*}_\text{max}$.
Fig.~\ref{fig:ductility_literature} shows that the theoretical deflection limit indeed provides a close upper bound on all the deflection results reported in the literature, supporting its validity.

\subsection{A general strategy to maximizing the deflection capacity of slab-like TIS} \label{sec:strategy}

\begin{figure}[H]
    \centering
    \includegraphics[width=0.45\linewidth]{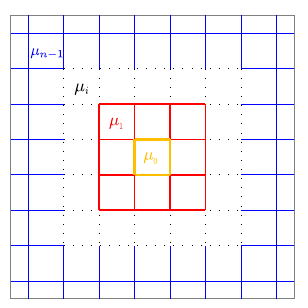}
    \caption{Friction coefficient design strategy for maximizing the deflection capacity of slab-like TIS. Each of the colored lines represents an interface with the indicated governing friction coefficient $\mu_i$.}
    \label{fig:varying_mu}
\end{figure}

The strategy for approaching the upper limit derived in Sec. \ref{sec:upper_bound} is based on creating a holistic global deformation mode which favors the engagement of all the blocks, particularly the more peripheral ones.
The common deformation mode of TIS slabs at large load levels is characterized by sliding of the central loaded block and large transverse displacements of it and its neighbors while the more peripheral blokcs undergo much smaller transverse displacements. 
Therefore, the key to reaching a global deformation mode engaging all the blocks is to prevent the premature/excessive sliding of the loaded block. 
We accomplish this by varying the friction coefficient $\mu$ on the interfaces in rings around the loaded central block, such that the values are highest at its interfaces (preventing its premature sliding) and gradually decrease farther away from the center, see Fig.~\ref{fig:varying_mu}. The strategy can be expressed in mathematical terms by the expression $\mu_0 \geq \mu_1 \geq ... \geq  \mu_{n-1}$, where $n = (N+1) / 2$ is the number of rings, or variation levels of the friction coefficient $\mu$, in a panel with $N$~x~$N$ internal blocks, see Fig. \ref{fig:varying_mu}.

In this study, we determined the values of the friction coefficients using a sequential algorithm. First, we determined an increased uniform base friction coefficient. Then, starting from outermost ring $\mu_{n-1}$ and continuing towards the center, we choose the $\mu$'s which maximizes the deflection ring by ring. This sequential approach not only follows the mechanical rationale of ensuring the participation of outer blocks by gradually reducing the friction coefficients in the outer rings, but it is also very efficient computationally compared with a multi-dimensional $\mu$ optimization, which would be an alternative option.
We note that, to allow a large range of $\mu$ variation across the rings (specifically assigning relatively low values for the peripheral rings), and to avoid premature sliding mechanisms, high friction coefficients are assigned to the internal rings, in particular for the interfaces around the central block. 
While such friction coefficients may be larger than those of common engineering materials, equivalent suppression of sliding can be obtained through geometrical means, e.g., by architecturing the surface morphology of the blocks \cite{Koureas2023,Djumas2017}\footnote{For this route of implementation, the expressions developed in Kouras et al. \cite{Koureas2023} linking the degree of required surface waviness needed to read a target friction coefficient, can be used.}. 
In that sense, the friction coefficients in our strategy should be interpreted as a general measure of interfacial sliding resistance rather than as a strictly bi-material interfacial/material property.

\subsection{The computational tool} \label{sec:LS-DEM}

The computational tool we use for the application of the strategy outlined in Sec. \ref{sec:strategy} is the Level-Set-Discrete-Element-Level (LS-DEM). This tool, originally developed for granular applications~\cite{Kawamoto2016}, has recently been adapted to structural analysis \cite{Feldfogel2022a}. It has been employed to model and study slab-like TIS \cite{Feldfogel2022, Feldfogel2023}, as well as the seismic behavior of multiblock tower structures \cite{Harmon2023}. Its main advantage is its computational efficiency and its ability to model the slip-governed response of TIS. More details on the application of LS-DEM to TIS can be found in \cite{Feldfogel2022,Feldfogel2023}. For completeness, we mention here the most salient assumptions, limitations, and modeling essentials:

\begin{itemize}[leftmargin=0.3cm]
    \item[-] The blocks experience contact and friction forces from adjacent blocks, as well as reaction, gravity, and damping forces. These forces determine their motion based on Newton's generalized law for three-dimensional rigid body dynamics.
    \item[-] Material non-linearity, specifically fracture, is not considered. This implies that the model is applicable only to scenarios where fracture of blocks does not influence the structural response appreciably.
    \item[-] A penalty approach is used to enforce contact and friction -- a linear law for the normal contact traction and a bilinear Coulomb’s law for the tangential friction traction, see \cite{Feldfogel2022}.
    \item[-] Based on the penalty-enforced contact, the closed-form expression $E = k_n \cdot \frac{L}{M}$ is used to link Young's modulus $E$ to the the penalty stiffness $k_n$, the number of internal interfaces $M$, and the side length of the panel $L$. The interested reader is referred to \cite{Feldfogel2022,Feldfogel2023} for details on this derivation.
\end{itemize}

\section{Results and discussion}
\label{sec:results}

\subsection{Examined set-up} \label{sec:setup}

\begin{figure}
    \centering
    \includegraphics[width=0.7\linewidth]{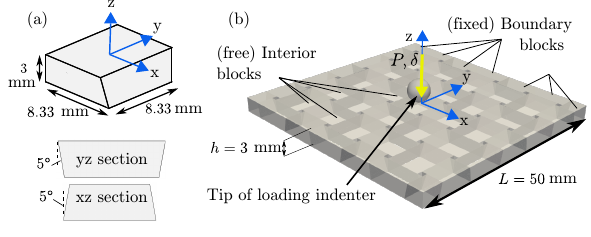}
    \caption{Examined slab-like TIS configuration with $5$ x $5$ internal blocks. (a) Typical internal block with cross-sections in xz and yz planes. (b) The panel's dimensions, boundary conditions, and loading. The Yellow arrow indicates the force P exerted by the indenter and corresponding indenter displacement $\delta$ in the negative $z$ direction}.
    \label{fig:configuration}
\end{figure}

The setup examined in this study is depicted in Fig.~\ref{fig:configuration}). It is based on a 5x5 panel made of truncated polyhedral block which was studied experimentally and numerically in \cite{Mirkhalaf2018, Feldfogel2022, Feldfogel2023}. 
The panel is subjected to a quasi-statically central indentation load applied with displacement control. The boundary conditions are fixed (i.e., zero displacements and rotations) peripheral blocks.
We chose this setup because (a) it represents a real TIS design which has been studied experimentally \cite{Mirkhalaf2018}, (b) the computational tool we used has been validated for it, see~\cite{Feldfogel2022}, and (c) it is representative of other TIS designs explored in the literature in terms of dimensions, material properties, number of blocks, and level of deflection, see Fig.~\ref{fig:ductility_literature}.

\subsection{Application of the strategy} \label{sec:application_of_strategy}

For the $5$ x $5$ reference set-up described in Fig.~\ref{fig:configuration}, $n=3$, and so the application of the general strategy outlined in Sec.~\ref{sec:strategy} comprises the following three steps: (1) Choosing a high uniform $\mu_0=\mu_1=\mu_2$; (2) Finding a range of optimal $\mu_2$, while keeping $\mu_0=\mu_1 \geq \mu_2$; and (3) Finding an optimal $\mu_1$, while $\mu_0 \geq \mu_1 \geq \mu_2$.

\begin{figure}[H]
    \includegraphics[width=1\textwidth]{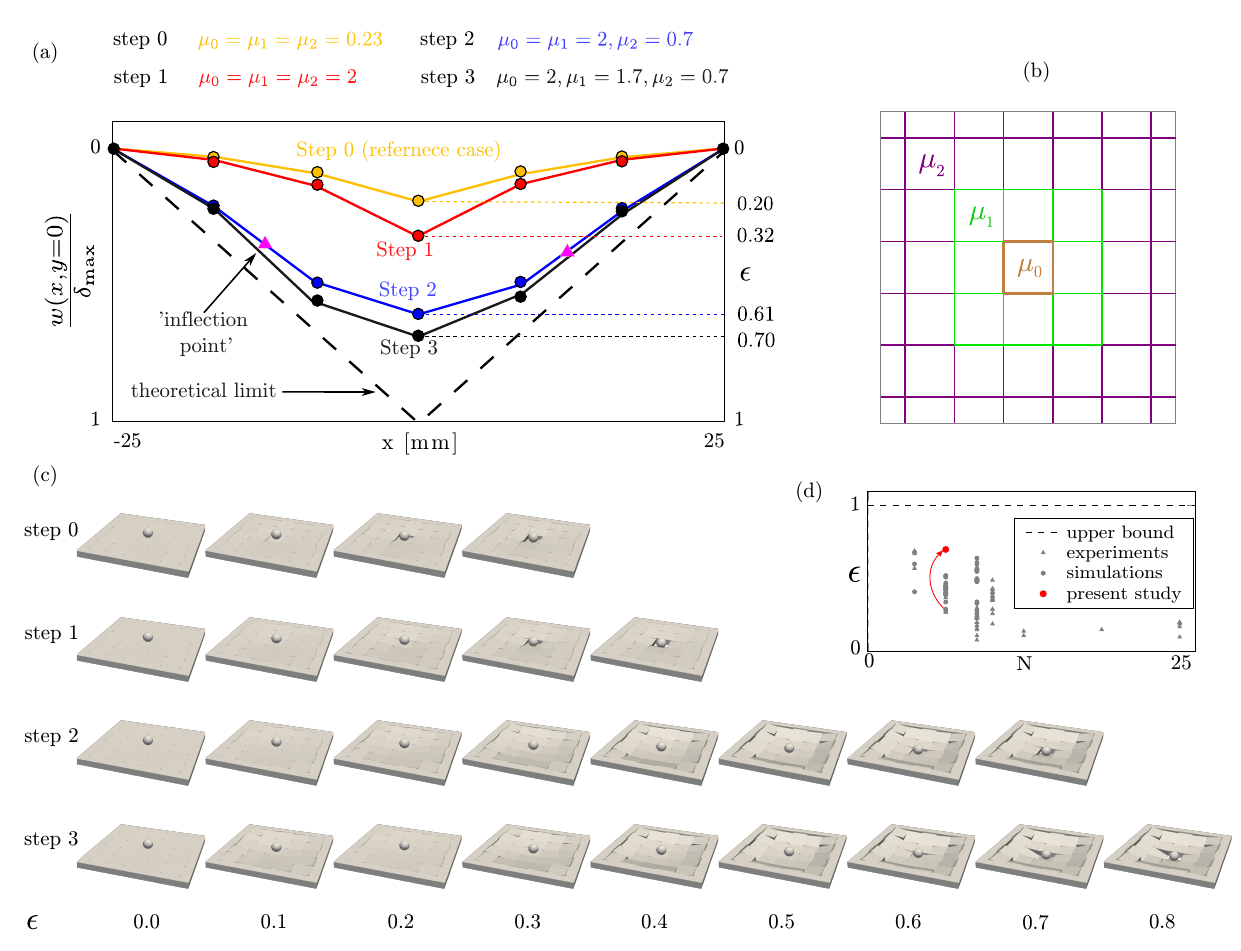} 
    \centering
    \caption{Illustration of the application of the strategy to increasing the deformation capacity of slab-like TIS panel. (a) Deformed shape of the central row in the $xz$-plane ($y=0$) corresponding to the reference configuration (step 0) and to the three steps of the strategy. Full circles represent the deflected positions of the centers of mass of the blocks and the lines between them are drawn as a visual aid. $w(x,y)$ denotes the displacement in the negative $z$ direction. The chosen friction coefficients for all the steps are given in the legend. The obtained deflection ratios in each of the four steps are listed on the right vertical axis. (b) A map indicating the interfaces associated with each $\mu$ ring. (c) Snapshots illustrating the evolution of the deformed configuration in each of the steps. (d) Comparison between the results reported in the literature and those obtained through the application of the proposed strategy.}
    \label{fig:application_of_strategy}
\end{figure}

The choice of $\mu$ in each step, the resulting increase in the deflection ratio, and the effects on the deformation response are depicted in Fig.~\ref{fig:application_of_strategy}.
For step 1, we chose a uniform $\mu_0=2.0$, as larger values had a negligible effect on increasing the deflection. 
Such saturation has been previously observed in \cite{Koureas2022,Koureas2023,Feldfogel2023}. 
Step 1 resulted in a moderate deflection increase from $\epsilon=0.20$ to $\epsilon=0.32$ in spite of the very large increase in $\mu$, see Fig. \ref{fig:application_of_strategy}(a).
The deformation mode remained fairly unchanged and governed by local sliding of the loaded block while the rest of the assembly remained fairly un-deformed, Fig. \ref{fig:application_of_strategy}(c).

Step 2, the first 'true' step of the strategy in the sense of introducing differences between the interfaces, resulted in major, almost two-fold, increase in deflection from $\epsilon=0.32$ to $\epsilon=0.61$, Fig. \ref{fig:application_of_strategy}(a).
In addition to this increase, we observe a qualitative shift in the deformation mode from a sliding-governed localized one in steps 0 and 1 to a global one resembling the deformation of monolithic slabs in step 2, Fig. \ref{fig:application_of_strategy}(c).
This is reflected both in the sequence of snapshots and in the 'inflection points' observed at the quarter-span in the central cross-sections in Fig.~\ref{fig:application_of_strategy}(a)\footnote{By 'inflection point' we mean the discrete equivalent of a true inflection point, that is, a change from decreasing slopes to increasing ones.}. Interestingly, a similarly positioned inflection points would occur in a monolithic slab under the same (fixed) boundary conditions, further supporting the similarity to monolithic equivalents. 
We also note that the rotation of the first rows of blocks is in accordance with the explanation of the chaining mechanism, as presented in Fig.~\ref{fig:chaining_effect}.
Step 3 leads to an increase of the deflection from $\epsilon=0.61$ to $\epsilon=0.70$ while retaining all the positive qualitative features of deformation observed for step 2.

In summary, the application of the strategy to the examined set-up has resulted in a 350$\%$ increase in the deflection factor from the reference $\epsilon=0.2$ to a value of $\epsilon=0.7$ -- the largest value hitherto found in the literature, see Fig.~\ref{fig:application_of_strategy}(d).
In fact, the nominal value we obtained in the simulations was $\epsilon=0.8$, making for a larger gap between the present ones and previously reported results. However, we discarded this results because, taking a conservative approach, we only considered the response up to the point where the central block lost contact with at least one of its neighbors (which occured at $\epsilon=0.7$).
From a qualitative standpoint, the application of the strategy resulted in a holistic global deformation mode with significant participation of all blocks, in accordance with our original intent.

\subsection{Practical significance} \label{sec:practical_importance}

\begin{figure}[H]
    \centering
    \includegraphics[width=1\linewidth]{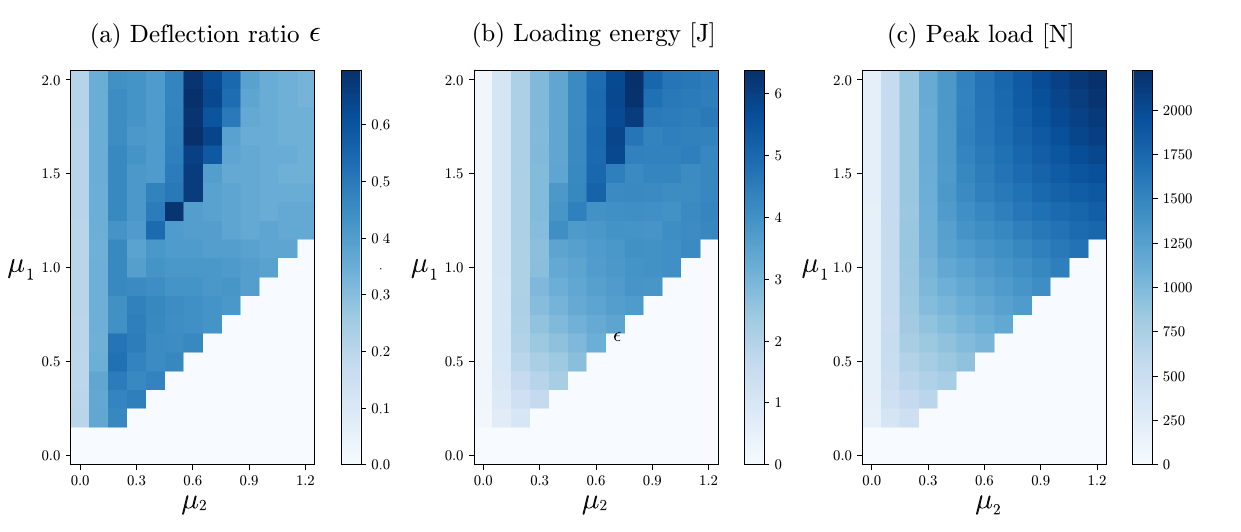}
    \caption{The variation of (a) deflection ratio $\epsilon$, (b) loading energy, and (c) peak load as a function of the friction coefficients $\mu_1 \geq \mu_2$ varied in rings according to Fig.~\ref{fig:varying_mu} with $\mu_0 = 2.0$.}
    \label{fig:matrix_plots}
\end{figure}

The main practical significance of the question of increasing the deflection capacity hinges on whether or not this increase is  associated with increased loading energy capacity, arguably the more important response parameter of the two. 
To investigate this point, we compare in Fig. \ref{fig:matrix_plots} the dependence of both, as well as of the peak load on the friction $\mu_1$ and $\mu_2$ for $\mu_0=2.0$, the values corresponding to steps 2 and 3. 
Starting with the peak load in Fig. \ref{fig:matrix_plots}(c) for comparative reasons, it increases monotonically with $\mu_1$ and $\mu_2$, see Fig.~\ref{fig:matrix_plots}(c), a result that is in line with previous results \cite{Feldfogel2023}.
Turning to the deflection and the loading energy in Fig.~\ref{fig:matrix_plots}(a,b), both display a very similar dependence on $\mu_1$ and $\mu_2$, follow a similar non-monotonic trend differently from the peak load, and peak together near $\mu_1 \approx 1.7$ and $\mu_2 \approx 0.6$. 
This similarity means that the strategy outlined in Sec.~\ref{sec:strategy} and illustrated in Sec.~\ref{sec:application_of_strategy} not only maximizes the deflection capacity of slab-like TIS but also, and perhaps more importantly, it maximizes their energy absorption/dissipation capacity.

\section{Conclusions}
\label{sec:conclusions}

In this study, we have discussed several previously unaddressed features of the deformation mechanics of slab-like TIS panels.
These include their unique load transfer mechanisms, the upper limit of their deflection capacity, and a systematic strategy to optimize this capacity and the commensurate energy absorption/dissipation capacity. 
We first described the 'chaining' load transfer mechanism, which sets apart slab-like TIS from beam-like ones and which allows them to develop much larger deflections.
Next, we proposed a theoretical limit of the deflection of slab-like TIS and showed that it provides a close upper-bound on all the results found in the literature.
We then illustrated the application of the proposed strategy to a reference slab and showed that it led to a holistic and global deformation, increased the deflection by 350$\%$, and resulted in the highest normalized deflection hitherto reported in the literature. 
Lastly, we showed that the proposed strategy not only significantly improves the deformation mode and deflection capacity, but that it also locally maximizes the energy dissipation/absorption capacity, one of TIS' salient advantages.
Future research avenues include an experimental application and validation of the proposed strategy, generalization and examination of the deformation mechanics under different load distributions, specifically uniformly distributed loads, and the possibility of using other design parameters in the strategy, e.g., by modifying the elastic modulus instead of the friction coefficient.

\section{Acknowledgement}
Shai Feldfogel was a Swiss Government Excellence Scholarship holder for the academic years 2021-2022 (ESKAS No. 2021.0165).
\printbibliography
\end{document}